\documentstyle[11pt,aaspp4]{article}

\input psfig

\font\FermiPPTfont=cmssbx10 scaled 1440
\font\FermiSmallfont=cmssq8 scaled 1200

\def\FNALppthead#1#2{
\null \vskip -1truein
\centerline{\hbox to 7.5truein {
\vbox to 1in{\vfill 
             \hbox{\psfig{figure=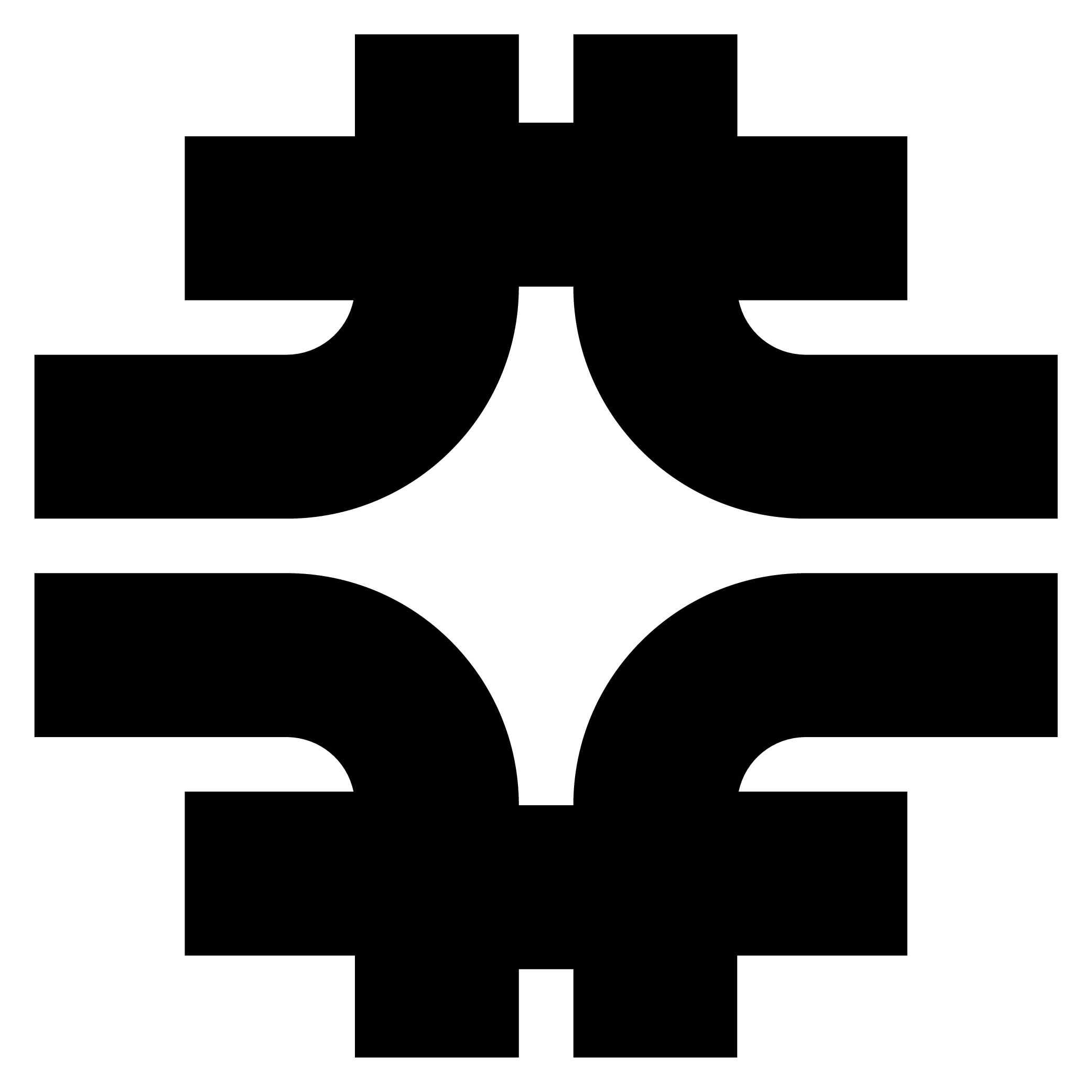,height=1.5cm,clip=}} 
             \vfill }
\hskip 1em
\vbox to 1in{\vfill
             \hbox{{\FermiPPTfont Fermi National Accelerator Laboratory}}
             \vfill}
\hfill
\vbox to 1in {\vfill \FermiSmallfont
              \hbox{#1}
              \hbox{#2}
              \vfill}
}}}

\def\s {\scriptscriptstyle}
\def\ccdot{{\hskip-0.7pt\cdot\hskip-0.7pt}}
\def\mathrelfun#1#2{\lower3.6pt\vbox{\baselineskip0pt\lineskip.9pt
  \ialign{$\mathsurround=0pt#1\hfil##\hfil$\crcr#2\crcr\sim\crcr}}}
\def\simlt{\mathrel{\mathpalette\mathrelfun <}}

\def\ln {{\rm ln}}
\def\sgn {{\rm sgn}}

\def\rme {{\rm e}}

\def\rmT {{\rm T}}
\def\bfp {{\bf p}}
\def\calO {{\cal O}}
\def\calH {{\cal H}}

\def\hatbfn  {{\hat{\bf n}}}
\def\pdbypd#1#2{{\partial#1\over\partial#2}}
\def\keV {{\rm \hbox{ke\kern-0.14em V}}}

\slugcomment{submitted to {\sl The Astrophysical Journal {\it Letters}}}

\lefthead{Stebbins}
\righthead{Extensions to Kompaneets \& S-Z}

\begin{document}

\FNALppthead{NASA/Fermilab Astrophysics Center}{Fermilab-Pub-97-???-A}

\title{Extensions to the Kompaneets Equation \\
       and Sunyaev-Zel'dovich Distortion}

\author{Albert Stebbins}
\affil{NASA/Fermilab Theoretical Astrophysics Group, Box 500, 
							Batavia, IL 60510} 

\begin{abstract}
Analytical expressions are presented for relativistic corrections to the
Kompaneets equation and the Sunyaev-Zel'dovich $y$-distortion. The latter
provides a convenient method of inferring both the temperature and Thomson
optical depth of the gas in clusters of galaxies using only observations of the
spectrum of CMBR photons passing through the cluster. The relativistic
correction gives an additional component of the S-Z effect apart from the
$y$-distortion and kinematic S-Z effect. The perturbative method used is shown
to provide a very accurate approximations when $kT_\rme\simlt100\,\keV$ but not
for higher temperatures.
\end{abstract}

  
\section{Introduction}

In the seminal paper, Zel'dovich and Sunyaev (1969), it is shown how
interactions of primordial photons with a hot plasma will lead to deviations
from a blackbody spectrum of photons.  The classical Sunyaev-Zel'dovich (S-Z)
effect gives the first order effect on the spectrum as one increases the
temperature of the electrons from zero and is not valid as the electron
velocities become relativistic.  In this {\it Letter} we give the next order
effect, which is important for hot gas in clusters of galaxies, describe a
perturbative expansion for computing higher order corrections, and consider the
convergence of this expansion.

The classical S-Z distortion was based on the Kompaneets equation
(\cite{Kompaneets57}) which describes how the spectrum of photons evolves under
the action of Compton scattering off of a stationary plasma of hot electrons.
To a first approximation, for non-relativistic electrons ($v^2\ll c^2$) and
soft photons ($\epsilon\ll m_\rme c^2$), kinematics dictates that Compton
scattering does not change the photon energy and there should be no change in
the photon energy distribution.  The Kompaneets equation describes the rate of
change in the photon spectrum to order ${v^2\over c^2}$ and ${\epsilon\over
m_\rme c^2}$. The next order term, $\propto{v^4\over c^4}$ or more precisely
$\propto\left({kT_\rme^2\over m_\rme c}\right)^2$, was presented in
\cite{Stebbins97a} and in this {\it Letter} we summarize these results while
adding other analytical results and applications. Compton scattering has been
well studied, by a variety of techniques and non-perturbative computations of
the spectral distortion of an incident blackbody spectrum passing through a hot
gas at arbitrary temperatures have been published (\cite{Fabbri81,Rephaeli95}).
The perturbative results of this {\it Letter} complement the more precise
numerical results in that they are analytical and describe the corrections to
the classical $y$-distortion in terms of a given spectral shape.  As we shall
see, this next order correction provides an good approximation to the numerical
results in the temperature regime relevant to the hot gas in clusters of
galaxies.  Rephaeli and Yankovitch (1997) have already shown that these
corrections, combined with relativistic corrections to the X-ray emission can
have significant effects on the estimates of the Hubble constant from cluster
observations.  A more expansive discussion of the perturbative results
presented here may be found in Stebbins~(1997a,b).

\section{Collisional Boltzmann Equation}

We describe the state of the primeval gas of photons in terms of the quantum 
mechanical occupation number in phase space, $n_\gamma$.  The evolution of
$n_\gamma$ can be described by the collisional Boltzmann equation which has the
form 
\begin{equation}
{D n_\gamma(\bfp_\gamma)\over Dt}=C(\bfp_\gamma)
\end{equation}
where $C(\bfp_\gamma)$ is the scattering term which describes the effect of 
scattering and ${D\over Dt}$ is a convective derivative along the photon's 
trajectory in phase space.   Assuming isotropy of the photon and 
electron distribution function, we need only solve for the change in $n_\gamma$
as a function of the photon energy, $\epsilon$, and not momentum, 
$\bfp_\gamma$.\footnote{If the photon are not at rest with respect to the 
electrons then their distribution functions will not both be isotropic in the 
same rest frame, which will lead to a {\it direction}-dependent spectral 
distortion known as the kinematic Sunyaev-Zel'dovich effect, which is a type of
Doppler shift.}  The collision term for Compton scattering may be written
(Stebbins~1997a,b)
\begin{eqnarray}
&&\hskip-13pt C(\epsilon,\Delta)
=\int d\beta\,f(\beta)\,\int d\Delta\,
\Biggl[{1\over(1+\Delta)^3}\overline{S}({\epsilon\over1+\Delta},\Delta)
         \,(1+n_\gamma(\epsilon))\,n_\gamma({\epsilon\over1+\Delta})        \cr
&&\hskip180pt -          \overline{S}(\epsilon,\Delta)
         \,(1+n_\gamma(\epsilon(1+\Delta)))\,n_\gamma(\epsilon)\Biggr]\ .
\label{EnergyCollisionIntegral}
\end{eqnarray}
where $\overline{S}(\epsilon,\Delta)$ gives an angle-average of the
differential cross-section:
\begin{equation}
\overline{S}(\epsilon,\Delta)=c\,N_\rme\,
\left\langle(1-\vec{\beta}\ccdot\hatbfn)\,{d\sigma\over d\Delta}\right\rangle
\ .
\label{Sbar}
\end{equation}
Here $N_\rme$ is the electron density, $\hatbfn$ is the incident electron
direction, $\vec{\beta}$ is the electron velocity in units of $c$, $f(\beta)$
gives the distribution of electron speeds, $\epsilon$ is the incident photon
energy, and $\Delta$ is gives the fractional change in the scattered photon,
i.e. $\epsilon'=\epsilon(1+\Delta)$.  The 1st term in the collision integral
gives the scattering into the beam and the 2nd term the scattering out of the
beam.

\section{Fokker-Planck Expansion}

As previously mentioned, for non-relativistic electrons and soft-photons,
kinematics dictates that the change in the photon energy is small so that 
the dependence of  $\overline{S}(\epsilon,\Delta)$ on it's second argument will
be very sharply peaked around $\Delta=0$.  The rest of the dependence on 
$\Delta$ in eq.~\ref{EnergyCollisionIntegral} is much smoother, and may be 
approximated by it's Taylor series about $\Delta=0$.  This Taylor expansion
yield a  Fokker-Planck type of approximation.  Taylor expanding the
$\Delta$-dependence of the distribution function $n_\gamma$  leaves only
derivatives of $n_\gamma$ evaluated at $\epsilon$.  These derivatives may be
taken out of the $\Delta$ and $\beta$ integrals and one is left with a
differential equation for $n_\gamma$ rather than an integro-differential
equation. The coefficients in the differential equation are determined by the
moments of the $\Delta$ distribution, which can be determined analytically when
the electron distribution is thermal.

The full Klein-Nishina cross-section is a rather complicated function, and for
the purposes of the CMBR we are mostly interested in the soft-photon
limit. This leads one to Taylor expand the full cross-section in powers of 
$\alpha={\epsilon\over m_\rme c^2}$ and at each order compute eq~\ref{Sbar}.
The 1st term is (Stebbins 1997a,b)
\begin{equation}
\overline{S}(\epsilon,\Delta)
=c\,N_\rme \sigma_\rmT\,\int_0^1 d\beta\,\beta\,
\overline{F}(\Delta,\beta\,\sgn(\Delta))+\calO(\alpha^1)
\label{SoftExpansion}
\end{equation}
where
\begin{eqnarray}
&&\overline{F}(\Delta,b)
=\sgn(\Delta)\times\calH(1-{(1-b)\,\Delta\over2b})\times\Biggl[
      {3(1-b^2)^2(3-b^2)(2+\Delta)\over16b^6}\,\ln{(1-b)(1+\Delta)\over1+b} \cr
&&\hskip-15pt
+{3(1-b^2)(2b-(1-b)\Delta)\over32b^6(1+\Delta)}\,
   (4(3-3b^2+b^4)+2(6+b-6b^2-b^3+2b^4)\Delta+(1-b^2)(1+b)\Delta^2)\Biggr]   \cr
&&\hskip200pt
\label{Fbar}
\end{eqnarray}
where $\calH$ is the Lorentz-Heaviside function.  With explicit functions such
as these it is tedious but straightforward to compute the Fokker-Planck
($\equiv$Taylor series) expansion of eq~\ref{EnergyCollisionIntegral} in powers
of $\alpha$ and $\beta$ or for a thermal electron distribution in powers of
$\Theta_\rme={kT_\rme\over m_\rme c^2}$.  Much of the methodology for such
expansions was developed in Barbosa (1982).

One may write this Fokker-Planck expansion as
\begin{equation}
{D n_\gamma\over D\tau}={1\over\epsilon^2}\pdbypd{}{\epsilon}\,\epsilon^3
\left[\sum_{n\ge0}\sum_{m\ge0}\Theta_\rme^n\alpha^m K^{\s(n,m)}[n_\gamma]
      \right] \qquad\Theta_\rme\equiv{kT_\rme\over m_\rme c^2} \qquad
                             \alpha={\epsilon\over m_\rme c^2}
\label{FokkerPlanckExpansion}
\end{equation}
where $K^{\s(n,m)}[n_\gamma]$ indicates a, potentially nonlinear, differential 
operator acting on $n_\gamma$ which does not depend on $m_\rme c^2$.  The 
pre-operator, ${1\over\epsilon^2}\pdbypd{}{\epsilon}$, is guaranteed by the 
form of eq~\ref{EnergyCollisionIntegral} which explicitly conserves photon 
number.  We have used the Thomson optical depth: $\tau=\int dt\,N_\rme 
\sigma_\rmT\,c$ where $\sigma_\rmT$ is the Thomson cross-section.  One finds 
that to include all terms that contribute to $K^{\s(n,m)}$ one must Taylor
expand to  order $2n+m$ in $\Delta$.  Since the energy redistribution is a
relativistic effect we find that $K^{\s(0,0)}=0$.  The classical Kompaneets
equation contains the two terms:
\begin{equation}
K^{\s(1,0)}[n_\gamma]=\epsilon\,\pdbypd{n_\gamma}{\epsilon} \qquad
K^{\s(0,1)}[n_\gamma]=(1+n_\gamma)\,n_\gamma\ .
\end{equation}
The 2nd order $\Delta$-expansion of Barbosa (1982) is only accurate enough to 
infer the one additional term
\begin{equation}
K^{\s(0,2)}[n_\gamma]=-{63\over5}(1+n_\gamma)\,n_\gamma\ .
\label{BarbosaInference}
\end{equation}
The new result (Stebbins97a,b) is
\begin{equation}
K^{\s(2,0)}= { 5\over 2}\epsilon  \,\pdbypd{  n_\gamma}{\epsilon  } 
            +{21\over 5}\epsilon^2\,\pdbypd{^2n_\gamma}{\epsilon^2}
            +{ 7\over10}\epsilon^3\,\pdbypd{^3n_\gamma}{\epsilon^3}
\label{KompaneetsExtension}
\end{equation}
which provides the next order correction for hotter electrons, but not for more
energetic photons.  Challinor \& Lasenby (1997) have confirmed these
corrections using an independent derivation and have also computed some
additional terms.  Combining this new terms with the classical result we find
an extended form of the Kompaneets equation:
\begin{equation}
\pdbypd{n_\gamma}{\tau}
={1\over \epsilon^2}\pdbypd{}{\epsilon}\,
\Biggl[\epsilon^3\Bigg({kT_\rme\over m_\rme c^2}\,
 \left(1+{5\over2}\Theta_\rme\right)\,\epsilon\,\pdbypd{n_\gamma}{\epsilon}
+{7\over10}\Theta_\rme^2\,
         \left( 6\epsilon^2\,\pdbypd{^2n_\gamma}{\epsilon^2}
               + \epsilon^3\,\pdbypd{^3n_\gamma}{\epsilon^3}\right)
+\alpha\,(1+n_\gamma)\,n_\gamma\Biggr)\Biggr]
\label{ExtendedKompaneets}
\end{equation}
which will be more accurate at higher electron temperatures than the classical
form.

\section{Extended Sunyaev-Zel'dovich Distortion}

The idea of the S-Z distortion is that one starts out with a background
radiation which is close to a blackbody spectrum, just what we expect to be
produced by the early universe, and it is {\it slightly} distorted by the
action of hot ionized gas through the Compton scattering process we have just
described.  In this small distortion limit we need just substitute a blackbody
spectrum with temperature $T_\gamma$ for $n_\gamma$ in the right-hand-side of
the Kompaneets equation and integrate over $\tau$ to obtain the distortion to
the spectrum.  If we apply this technique to the Fokker-Planck expansion we
obtain a series for the total distortion:
\begin{equation}
\Delta n_\gamma=
  \sum_{n\ge0}\sum_{m\ge0}Y_{\s\rm C}^{\s(n,m)}\Delta n_{\s\rm SZ}^{\s(n,m)}
\qquad Y_{\s\rm C}^{\s(n,m)}=\int d\tau\,\Theta_\rme^n\,\Theta_\gamma^m 
\qquad \Theta_\gamma\equiv{kT_\gamma\over m_\rme c^2}
\label{SZexpansion}
\end{equation}
where
\begin{equation}
\Delta n_{\s\rm SZ}^{\s(n,m)}(x)=K^{\s(n,m)}[n^{\s\rm BB}_\gamma] \qquad
n^{\s\rm BB}_\gamma={1\over e^x-1} \qquad   x={\epsilon\over k T_\gamma}\ .
\end{equation}
From the known $K^{\s(n,m)}$ one finds
\begin{eqnarray}
&&\hskip-33pt
\Delta n_{\s\rm SZ}^{\s(0,0)}(x)=0                                          \cr
&&\hskip-33pt
\Delta n_{\s\rm SZ}^{\s(1,0)}(x)=-\Delta n_{\s\rm SZ}^{\s(1,0)}(x)
               ={5\over63}\Delta n_{\s\rm SZ}^{\s(2,0)}(x)
               ={x e^x\over(e^x-1)^2}\,\left(x\,{e^x+1\over e^x-1}-4\right) \cr
&&\hskip-33pt
\Delta n_{\s\rm SZ}^{\s(2,0)}(x)={x e^x\over(e^x-1)^2}\,
        \left(- 10
              +{47\over 2}x\,{ e^x+1                 \over e^x-1   }
              -{42\over 5}x^2{        e^{2x}+ 4e^x+1 \over(e^x-1)^2}
               +{ 7\over10}x^3{(e^x+1)(e^{2x}+10e^x+1)\over(e^x-1)^3}\right)\cr
&&\hskip200pt
\label{ExtendedYdistortion}
\end{eqnarray}
Since one expects that to each order in energy that a blackbody spectrum is a
stationary solution when the electron and photon temperature are equal, one
expects the sum rule
\begin{equation}
\sum_{n=0}^N \Delta n_{\s\rm SZ}^{\s(n,N-n)}(x)=0 \ .
\end{equation}
This sum rule tell us that between the $N+1$ terms in this sum only $N$ give
linearly {\it independent} spectral shapes for the distortion.  Applying the
sum rule to $N=2$ one can infer the function
$\Delta n_{\s\rm SZ}^{\s(1,1)}(x)$ via the relation
$\Delta n_{\s\rm SZ}^{\s(1,1)}=-\Delta n_{\s\rm SZ}^{\s(2,0)}
-\Delta n_{\s\rm SZ}^{\s(0,2)}(x)$. 

The classical S-Z $y$-distortion includes only $\Delta n_{\s\rm SZ}^{\s(1,0)}$
and $\Delta n_{\s\rm SZ}^{\s(0,1)}$ and is parameterized by the
``$y$-parameter'':
\begin{equation}
y=Y_{\s\rm C}^{\s(1,0)}-Y_{\s\rm C}^{\s(0,1)}
=\int d\tau\,{k(T_\rme-T_\gamma)\over m_\rme c^2}\ .
\label{Yparameter}
\end{equation}
Additional terms in the Fokker-Planck expansion introduces additional 
parameters such as $Y_{\s(2,0)}$ which are measures of different moments of the
electron or photon temperature.

\section{Accuracy and Convergence}

\begin{figure}
\plotone{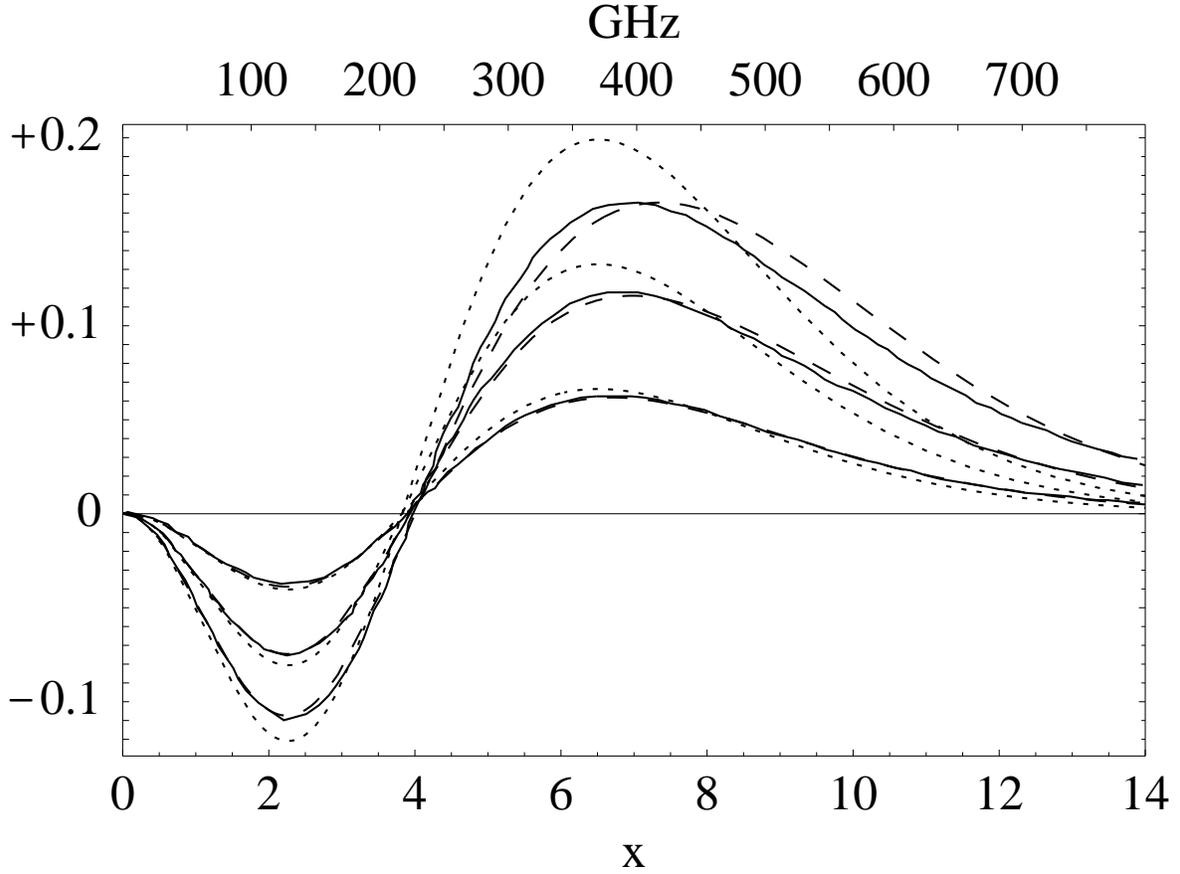}
\caption{Plotted vs. frequency is the shape of the spectral distortion ($x^3
\Delta n(x)\propto$\,brightness) to thermal CMBR photons passing through hot
gas.  Shown are three triplets of curves for gas temperatures of 5, 10, and
$15\,\keV$ in order of increasing amplitude.  In each triplet the dotted curve
shows the classical $y$-distortion, the dashed curve the $\calO(\Theta_\rme^2)$
perturbative prediction, and the solid curve the ``true'' distortion taken from
the numerical results of Rephaeli~(1995).}
\label{fig:Comparison}
\end{figure}

Expansions like those of eq.s~\ref{FokkerPlanckExpansion}\&\ref{SZexpansion}
are most useful when good accuracy is obtained with only the first few
terms. These expansions give the correct Taylor series expansion in
$\Theta_\rme$ and $\Theta_\gamma$ about zero, but this may not yield a good
approximation even when these variables are small, and in fact these series may
not converge for any value of the parameters.  As a first check we have
compared the series expansion of $\Delta n_\gamma$ to order $\Theta_\gamma^0$
and $\Theta_\rme^2$ to the numerical results of Rephaeli~(1995) in
fig~\ref{fig:Comparison} for $\alpha=0$ and $kT_\rme=5\,\keV,\ 10\,\keV\
,15\,\keV$ ($\Theta_\rme=0.0098,\ 0.0196$, and $0.0294$). The truncated series
yields a good approximation to the numerical results, but with discernible
differences. While it is perhaps surprising that such large corrections to the
classical S-Z distortion are present at such low temperatures, the fact that
the next order correction makes up for most of the total correction indicates
that this series will be useful.

\begin{figure}
\plotone{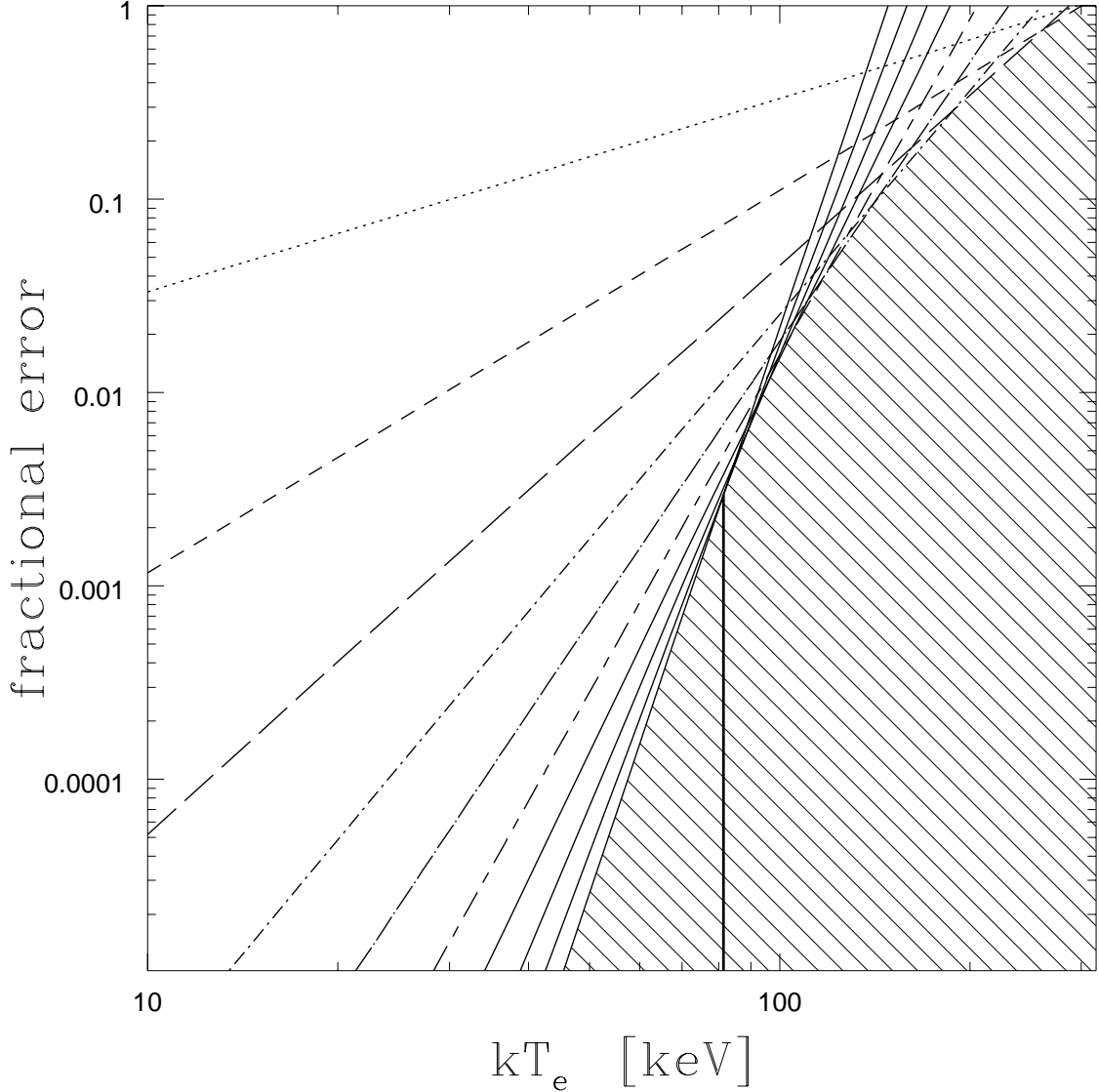}
\caption{Plotted versus electron temperature is the fractional error made by
truncated Taylor series approximations to the Rayleigh-Jeans spectral
distortion.  The curves are: ({\it dotted}) the classical S-Z effect - a
$\calO(\Theta_\rme^1)$ truncation; ({\it short-dashed}) a
$\calO(\Theta_\rme^2)$ truncation; ({\it long-dashed}) a $\calO(\Theta_\rme^3)$
truncation , and so on up to an $\calO(\Theta^{10})$ truncation.  The hashed
region to the right of the vertical line, and some of the hashed region to its
left are accuracies which are not attainable at any order in the Taylor series.
Some of the hashed region to the left of the vertical line is attainable by
truncating at orders greater than $\calO(\Theta_\rme^{10})$.  Note that while
the Taylor series is never very accurate when $kT_\rme>100\,\keV$, a truncated
Taylor series can be extremely accurate at lower temperatures.}
\label{fig:ErrorEnvelope}
\end{figure}

	To examine the accuracy and convergence of the series expansions
one may look at the low frequency limit of the S-Z distortion. By substituting
the power series expansion in $x$ of the blackbody spectrum into
eq~\ref{EnergyCollisionIntegral} one obtains a power series expansion for the
extended S-Z distortion:
\begin{equation}
\Delta n^{\s\rm SZ}_\gamma(x)=r_{-1}(\Theta_\rme)\,{1\over x}
          +r_1(\Theta_\rme)\,x+r_3(\Theta_\rme)\,x^3+\calO(x^5,\Theta_\gamma^1)
\end{equation}
the even powers having zero coefficients.  Here we will be interested in the
``Rayleigh-Jeans'' term, $r_{-1}(\Theta_\rme)$, which gives the fractional
change in brightness at the lowest frequencies.  This is given by
\begin{eqnarray}
&&\hskip-20pt r_{-1}(\Theta_\rme)=
  \int_0^1 d\beta\,{\beta^2\gamma^5\,e^{-\gamma/\Theta_\rme}
                    \over\Theta_\rme K_2({1\over\Theta_\rme})}
  \Biggl[{9-15\,\beta^2+10\beta^4\over4\beta^4}
  -{3\,(3-6\beta^2+4\beta^4-\beta^6)\over4\beta^5}\,\ln{1+\beta\over1-\beta}\cr
&&\hskip130pt
  -{3(3-\beta^2)(1-\beta^2)^2\over16\beta^6}\,
                                 \left(\ln{1+\beta\over1-\beta}\right)^2\Biggr]
                                                        +\calO(\Theta_\gamma^1)
\end{eqnarray}
where $\gamma=(1-\beta^2)^{-{1\over2}}$ and $K_2$ is a modified Bessel
function. While one may not be able to express this integral in terms of simple
functions it is easily evaluated numerically. One may also compute the Taylor
series of $r_{-1}(\Theta_\rme)$ about zero:
\begin{equation}
r_{-1}(\Theta_\rme)=-           2             \Theta_\rme 
                    +{         17\over      5}\Theta_\rme^2
                    -{        123\over     20}\Theta_\rme^3
                    +{       1989\over    140}\Theta_\rme^4
                    -{      14403\over    320}\Theta_\rme^5
                    +{      20157\over    112}\Theta_\rme^6
   +\calO(\Theta_\rme^7,x^1,\Theta_\gamma^1).
\end{equation}
which also gives the low frequency limit of the expansion of
eq~\ref{SZexpansion} with $m=0$. The convergence of this series should tell us
something about the convergence of eq~\ref{SZexpansion}.  In
fig~\ref{fig:ErrorEnvelope} is plotted the fractional error of truncated Taylor
series approximations to $r_{-1}$.  It seems clear that the series does not
converge for $kT_\rme>80\,\keV$ and likely that it has zero radius of
convergence. However below $\sim100\,\keV$ the 1st few terms in the Taylor
series do provide excellent approximations to $r_{-1}$.  If we take this to be
representative of the more general series of eq~\ref{SZexpansion} we may
conclude that these extensions to the S-Z distortion should be sufficient for
application to clusters of galaxies where we expect $kT_\rme\simlt20\keV$.  One
just need be careful not to expand to such larger orders that the series starts
to diverge.\footnote{These convergence properties are indicative of an {\it
asymptotic} series, an example of which is perturbation theory in quantum field
theory.}

It is not clear whether one might be able to slightly modify
eq~\ref{SZexpansion} to obtain a series which is useful at all temperatures.
We know that for ultra-relativistic electrons that essentially every scattering
will take a photon out of the Rayleigh-Jeans region to much higher
energies. Thus we expect the brightness at low frequencies to be degraded to a
gray-body spectrum, the brightness decreased by the factor $e^{-\tau}$. This
reasoning correctly predicts the limit $r_{-1}(+\infty)=-1$. This asymptotic
behaviour makes the error of any approximant which is a polynomial in
$\Theta_\rme$ diverge at large $\Theta_\rme$.  Once can hope to improve
convergence if one instead Taylor expands in some other variable
$X=f(\Theta_\rme)$ where $f(0)=0$ and $f(\infty)$ is finite.  For example if
one takes $X=\Theta_\rme/(1+3\Theta_\rme)$ the Taylor series of $r_{-1}$ in
$X$, 
\begin{equation}
r_{-1}(\Theta_\rme)=-            2              X 
                    -{          13\over       5}X^2
                    -{          15\over       4}X^3
                    -{         117\over      35}X^4
                    -{        3189\over    2240}X^5
                    +{        6261\over     448}X^6
                    +{       87117\over    2560}X^7+\calO(X^{11})\ ,
\end{equation}
seems to have good convergence for {\it all} temperatures.  Note that the
Taylor series in $X$ contains no more information than that in $\Theta_\rme$,
since one can derive one from the other. What is unclear is whether one can
make something like this work for all frequencies.

	At cluster temperatures the accuracy obtained from the 1st few terms
in eq~\ref{SZexpansion} is so good that other effects may be the leading source
of errors.  One is the finite optical depth effect.  If one expands in $\tau$
as well as in temperature the leading $\tau^2$ term is that given in
Fabbri~(1981)  which is $\sim\calO(\tau^2\Theta_\rme^2)$.  For clusters, where
$\tau\sim0.01$, this is liable to be much smaller than the $Y_{\s(2,0)}\Delta
n^{\s\rm SZ}_{\s(2,0)}$ correction given above but should be roughly the same
order as the next term, $Y_{\s(3,0)}\Delta n^{\s\rm SZ}_{\s(3,0)}$,
which we have not computed.  To properly compute finite optical depth
corrections in a non-uniform medium (such as a cluster) one must really do a
proper radiative transfer computation.

\section{Application to Clusters}

Given that only the first few terms in eq~\ref{SZexpansion} give an excellent
approximation to the exact result for cluster temperatures, we may think of the
extended S-Z distortion of the CMBR spectrum as consisting a few distinct types
of distortions corresponding to the different $n$ and $m$ in that equation.
For the CMBR passing through clusters the terms with $m\ne0$ are 
negligible and we henceforth ignore them.  Of the remaining terms the largest
is $\Delta n^{\s\rm SZ}_{\s(1,0)}$ (the classical S-Z effect) the 2nd largest
is $\Delta n^{\s\rm SZ}_{\s(2,0)}$, etc. The amplitude of the classical S-Z
distortion is given by $y=Y_{\s(1,0)}$ and tells one about the product of the
temperature and the optical depth.  The amplitude of the next term,
$Y_{\s(2,0)}$, can be combined with $Y_{\s(1,0)}$ to give a weighted measure of
the temperature and optical depth separately
\begin{equation}
\overline{kT_\rme}=m_\rme c^2\,{Y_{\s\rm C}^{\s(2,0)}\over
                                Y_{\s\rm C}^{\s(1,0)}}
={\int d\tau\,(kT_\rme)^2\over\int d\tau\,kT_\rme} \qquad
\overline{\tau}={{Y_{\s\rm C}^{\s(1,0)}}^2\over Y_{\s\rm C}^{\s(2,0)}}
         ={\left[\int d\tau\,(kT_\rme)\right]^2\over\int d\tau\,(kT_\rme)^2}\ .
\label{Ttau}
\end{equation}
If the electron temperature was constant along a line-of-sight, then we expect
each of the ratios, ${Y^{\s(n+1,0)}\over Y^{\s(n,0)}}$, to be the same
independent of $n$. However one can use deviations from this rule to detect
non-isothermality.   For gas as hot as $15\,\keV$ ${Y_{\s(2,0)}\over
Y_{\s(1,0)}}\sim0.03$  but since in the 100-400\,GHz region ${\Delta n^{\s\rm
SZ}_{\s(2,0)}\over \Delta n^{\s\rm SZ}_{\s(1,0)}}\sim 5$ the signal from the
next order correction to the classical $y$-distortion is only about 7 times
smaller than the classical effect.  For many hot clusters one can expect
this next order effect to be at the ${\Delta T\over T}\sim10^{-5}$ level which
is certainly accessible with existing telescopes and should also be measurable
by the all-sky survey of the PLANCK satellite.

\begin{figure}
\plotone{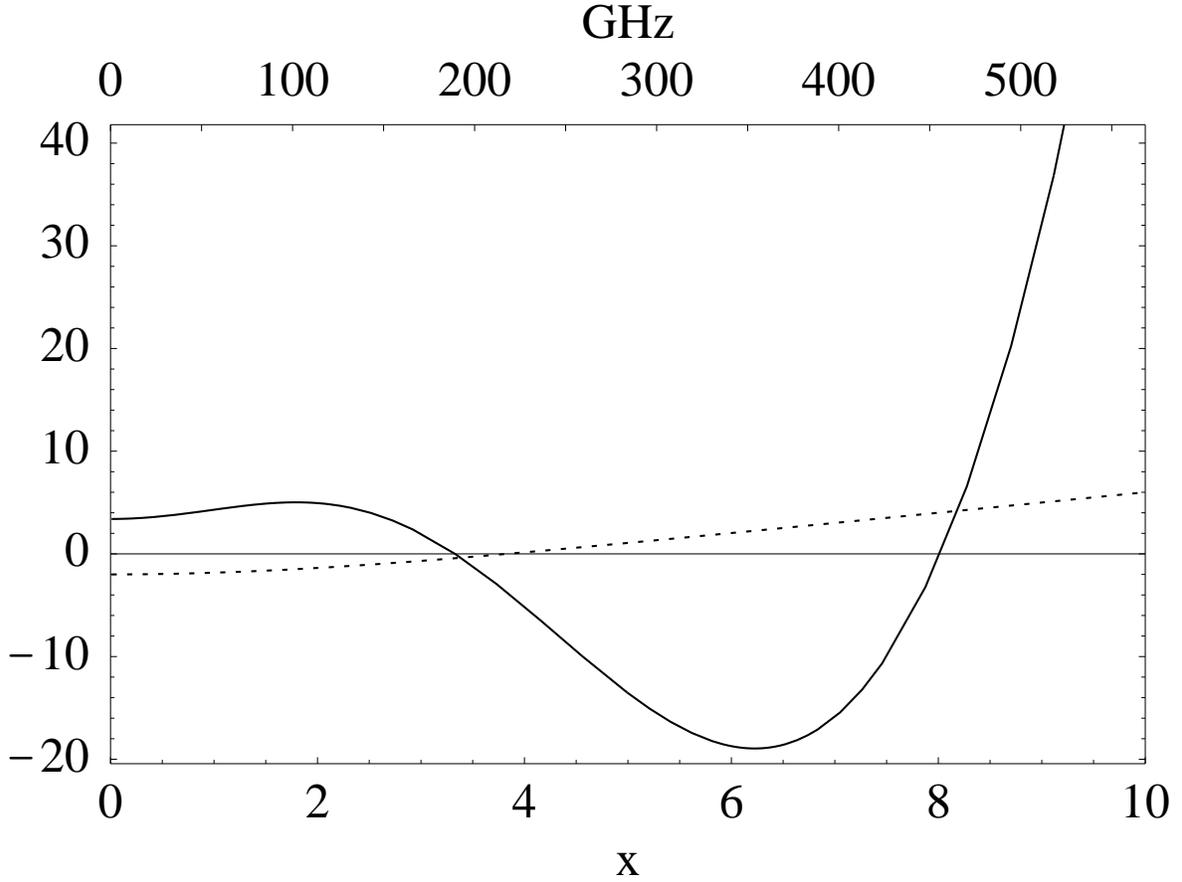}
\caption{Plotted versus frequency is ${\Delta T\over T}$ from the classical
$y$-distortion: $\Delta n^{\s\rm SZ}_{\s(1,0)}$ ({\it dotted line}), and from 
the next order correction: $\Delta n^{\s\rm SZ}_{\s(2,0)}$ ({\it solid line}).
The fact that the solid curve has greater amplitude than the dotted curve
indicates that ${Y_{\s(2,0)}\over Y_{\s(1,0)}}\sim {kT_\rme\over m_\rme c^2}$
underestimates the size of this correction.  The distinct shape of the
correction should allow observations with high frequency coverage to separate
out this component of the distortion from other sources of foreground
contamination.}
\label{fig:shape}
\end{figure}

In order to measure quantities like $Y_{\s(2,0)}$ one must be able to
distinguish the $\Delta n^{\s\rm SZ}_{\s(2,0)}$ distortion from other effects
such as anisotropies, the classical $y$-distortion, and contamination by dust,
synchrotron, and free-free emission.  In fig~\ref{fig:shape} we plot the shape
of $\Delta n^{\s\rm SZ}_{\s(2,0)}(x)$. One sees that it's shape is quite
different than any of the other distortions just mentioned as long as one looks
over a sufficient frequency range, say from $100-400\,$GHz.

Historically the frequency at which the total S-Z distortion to the spectrum is
zero has been of great interest.  From the eq~\ref{SZexpansion} one may compute
the Taylor series describing how this frequency changes as the electron
temperature increases.  Using $\Delta n^{\s\rm SZ}_{\s(1,0)}(x)$ and $\Delta
n^{\s\rm SZ}_{\s(2,0)}(x)$ one may determine the linear term,
\begin{equation}
x_0(\Theta_\rme)=3.83002+4.29189{Y_{\s(2,0)}\over Y_{\s(1,0)}}
                                                          +\calO(\Theta_\rme^2)
\end{equation}
which agrees with Fabbri (1981).  Rephaeli (1995) has pointed out that this
linear extrapolation overestimates the shift in $x_0$ which suggests that the
next term in the series is negative and this is indeed found by Challinor \&
Lasenby (1997).

Traditionally X-ray measurements are used to determine the gas temperature in
galaxy clusters. Quantities like those in eq~\ref{Ttau} are complementary
measures because the X-rays give a temperature weighted by the square of the
electron density while the S-Z effect is weighted linearly with the electron
density.  Thus X-ray temperatures are more weighted toward the cluster centers
than S-Z based temperatures.  Combining X-ray measurements with S-Z
measurements and extensions will allow better empirically based modeling of
clusters.  More generally one can hope to make temperature determinations using
eq~\ref{Ttau} in gas clouds which are too diffuse to produce sufficient X-ray
luminosity for a good temperature determination.  Hubble constant
determinations must rely on X-ray measurements since here one makes use of the
different density dependencies of X-ray emission and S-Z distortion.  One might
consider looking for the finite optical depth effects which do depend on a
different power of the density, but these are liable to be very difficult to
measure (\cite{Fabbri81}).

	In summary, it is proposed that one fit for an additional component of
the CMBR spectral distortion in the direction of galaxy clusters or other hot
gas where a $y$-distortion has been detected.  The amplitude of this additional
component, $Y_{\s(2,0)}$, may be used to gain additional information about the
gas temperature and density.

\acknowledgments
Special thanks to Charley Lineweaver and Jim Bartlett for encouraging me in
this work. This work was supported by the DOE at Fermilab and by the NASA grant
5-2788.

\end{document}